\documentclass[english]{article}
\usepackage[T1]{fontenc}
\usepackage[latin9]{inputenc}
\usepackage{array}
\usepackage{refstyle}
\usepackage{units}
\usepackage{amstext}
\usepackage{graphicx}
\usepackage{wasysym}

\makeatletter

\AtBeginDocument{\providecommand\secref[1]{\ref{sec:#1}}}
\AtBeginDocument{\providecommand\subsecref[1]{\ref{subsec:#1}}}
\AtBeginDocument{\providecommand\figref[1]{\ref{fig:#1}}}
\AtBeginDocument{\providecommand\tabref[1]{\ref{tab:#1}}}
\providecommand{\tabularnewline}{\\}
\RS@ifundefined{subsecref}
  {\newref{subsec}{name = \RSsectxt}}
  {}
\RS@ifundefined{thmref}
  {\def\RSthmtxt{theorem~}\newref{thm}{name = \RSthmtxt}}
  {}
\RS@ifundefined{lemref}
  {\def\RSlemtxt{lemma~}\newref{lem}{name = \RSlemtxt}}
  {}

\newcommand{\lyxaddress}[1]{
\par {\raggedright #1
\vspace{1.4em}
\noindent\par}
}

\@ifundefined{showcaptionsetup}{}{%
 \PassOptionsToPackage{caption=false}{subfig}}
\usepackage{subfig}
\makeatother

\usepackage{babel}
\begin{document}

\title{Impact of Time Of Flight on early-stopped Maximum-Likelihood Expectation
Maximization PET reconstruction}

\author{L.Presotto\textsuperscript{1}, E De Bernardi\textsuperscript{2}
and V Bettinardi}
\maketitle

\lyxaddress{{\large{}}\textsuperscript{{\large{}1}}{\large{}Nuclear Medicine
Unit, IRCCS Ospedale San Raffaele, Italy }\textsuperscript{{\large{}2}}{\large{}Medicine
and Surgery Department, Milano Bicocca University, Milano, Italy}}
\begin{abstract}
The use of Time Of Flight (TOF) in Positron Emission Tomography (PET)
is expected to reduce noise on images, thanks to the additional information.
In clinical routine, a common reconstruction approach is the use of
maximum likelihood expectation maximization (MLEM) stopped after few
iterations. Empirically it was reported that, at matched number of
iterations, the introduction of TOF increases noise. In this work
we revise the theory describing the signal and noise convergence in
MLEM, and we adapt it to describe the TOF impact on early stopped
MLEM. We validated theoretical results using both computer simulations
and phantom measurements, performed on scanners with different coincidence
timing resolutions. This work provides theoretical support for the
empirically observed noise increase introduced by TOF. Conversely,
it shows that TOF not only improves signal convergence but also makes
it less dependent on the activity distribution in the field of view.
We then propose a strategy to determine stopping criteria for TOF-MLEM,
which reduces the number of iterations by a factor proportional to
the coincidence timing resolution. We prove that this criteria succeeds
in markedly reducing noise, while improving signal recovery robustness
as it provides a level of contrast recovery which is independent from
the object dimension and from the activity distribution of the background.
\end{abstract}

\section{Introduction}

Time of flight (TOF) positron emission tomography (PET) provides additional
information in the acquired data respect to conventional non-TOF PET.
On top of the sufficient tomographic transform of the data, it provides
a range of probable emission locations . The tomographic problem consequently
results better conditioned and higher signal to noise ratios (SNR)
can be obtained on reconstructed images. In a TOF system characterized
by a coincidence timing resolution (CTR) $\Delta t$, expressed in
$ps$ full width half maximum (FWHM), $D_{eff}\equiv\sqrt{2\pi}\frac{\Delta t}{\sqrt{8\ln2}}\nicefrac{c}{2}$
is the ``effective diameter'' of the TOF kernel. In the central
pixel of a uniform circular object of diameter $D$, TOF is expected
to provide a noise reduction of a factor $\sqrt{\nicefrac{D}{D_{eff}}}$
\cite{Vunckx2010}. 

In clinical setups a standard way to reconstruct images is Maximum
Likelihood Expectation Maximization (MLEM)\cite{Shepp1982}, stopped
quite far from convergence to achieve a noise reduction, exploiting
the difference in convergence speed for signal and for noise. Most
often MLEM is used in combination with the ordered subset acceleration,
in the so called OSEM algorithm\cite{Hudson1994}, however it is assumed
that, especially in the early iterations, the use of the OSEM acceleration
does not influence the results. It has already been observed that
TOF increases MLEM convergence speed, therefore both signal and noise
are higher at matched number of iterations\cite{Bettinardi2011,Conti2011a}.
An improved signal to noise ratio could be therefore achieved by simply
modifying the early stopping criteria to either increase contrast
or reduce noise. It was already reported that to achieve an improvement
in image quality when TOF is present, a lower number of iterations
is recommended \cite{Conti2013}. Such approaches were based on the
concept of ``SNR matching'': achieving the same SNR with a reduced
counting statistics or using variations of the same concept (e.g.:
constant noise and better contrast recovery, lower noise at identical
signal recovery etc...). The limit of these approaches is that it
is not clear how to find the number of iterations that satisfies such
a request. As an example, in a study the noise level was measured
as the standard deviation between neighboring voxels in a region of
interest over the liver, then the selecting the patient-specific number
of iteration for that specific patient that provided equal levels
of noise than a non-TOF reconstruction\cite{Karp2008}. Similarly,
matching contrast recovery on phantom images might be misleading as
convergence speed depends on many factors, and therefore can change
from acquisition to acquisition.

While the need to modify the stopping criteria when using TOF is clear,
especially from empirical observations, theoretical rules to guide
the choice of the number of iterations in an a-priori way are lacking.
The aim of this study is to theoretically derive and experimentally
validate the convergence of signal and noise in non-TOF and in TOF
MLEM with different CTRs in the range of iterations used in clinics
(i.e. far from convergence). This allows optimal exploitation of the
potentialities offered by new TOF-PET systems, with CTR as low as
$210\;ps$\cite{VanSluis2019}. The final aim is to develop an easy
stopping criterion for TOF MLEM. It should be able to generate images
with a minimal level of noise, while retaining quantitative accuracy
and improving robustness to changes in the composition of the objects
within the FOV. 

This paper is structured as follows. After notations and conventions,
in \secref{Theory} we analyze the convergence speed of signal and
noise in TOF and non-TOF MLEM. We review the theory of Barrett et
al \cite{Barrett1994} about noise as a function of iterations and
extend it to TOF. In \secref{Simulations} and \ref{sec:Phantom-measurements}
we perform computer simulations and phantom measurements for theory
validation. In \secref{Proposed-early-stopping}, before discussion,
we propose an iteration stopping criteria optimized for TOF.

\paragraph{Notations and conventions}

We indicate with $\lambda$ image voxel activity values, with $y$
sinogram counts and with $c_{i,j}$ the probability that a photon
pair emitted from pixel $j$ is recorded in the sinogram bin $i$.
We use $k$ to indicate the iteration number, using it as a superscript
when referring to the estimate of a parameter at the $k$-th iteration.
To avoid confusion with exponentiation, we indicate the latter with
parenthesis; e.g.: $\lambda^{k}$ is the estimate of $\lambda$ at
the $k$-th iteration while $\left(\alpha\right)^{k}$ is the $k$-th
power of $\alpha$. The forward projection operator is indicated in
component notation as $y_{i}=\sum_{j}c_{i,j}\lambda_{j}$, and in
matrix notation as $y=H\left[\lambda\right]$, while the backprojector
is indicated with $H^{T}$. We define the normalization factor $\eta_{j}=\sum_{i}c_{i,j}$.
Matrix multiplications will be highlighted by the use of square brackets
(e.g.:$y=H\left[\lambda\right]$) while element-wise multiplications
will be identified by the Hadamard product ``$\circ$'' . Therefore,
the forward projection of the element-wise multiplication between
$\lambda$ and a matrix $\delta$ will be written as $H\left[\lambda\circ\delta\right]$.
Element-wise divisions are written as standard fractions. Note that
throughout the paper we will not use resolution modeling within the
$H$ operators. Point spread function modeling indeed makes unconstrained
MLEM reconstruction undetermined at high frequencies\cite{Nuyts2014},
therefore studying the convergence of both signal and noise at those
frequencies is not possible, as it is dependent on the specific implementation.

\section{Theory\label{sec:Theory}}

\subsection{Objects convergence\label{subsec:Objects-convergence-theory}}

We start analyzing MLEM convergence proprieties. In PET reconstruction
the tomographic problem is the minimization of the negative likelihood
\begin{equation}
\lambda^{*}=\arg\min_{\lambda}\sum_{i}\bar{y_{i}}-y_{i}\ln\bar{y_{i}}
\end{equation}
 with $\bar{y_{i}}=\sum_{j}c_{i,j}\lambda_{j}$. It has been shown
that MLEM can be also written as a gradient descent algorithm with
unitary step size and a diagonal preconditioner equal to the current
image estimate divided by the normalization matrix\cite{Lange1990}.
Mathematically 
\begin{equation}
\lambda_{j}^{k+1}=\lambda_{j}^{k}+\nicefrac{\lambda_{j}^{k}}{\eta_{j}}\sum_{i}c_{i,j}\frac{y_{i}-\sum_{\xi}c_{i,\xi}\lambda_{\xi}}{\sum_{\xi}c_{i,\xi}\lambda_{\xi}}\label{eq:MLEM_PGD}
\end{equation}
or, in matrix notation, 
\begin{equation}
\lambda^{k+1}=\lambda^{k}+M\left[H^{T}\left[W\left[y-H\lambda^{k}\right]\right]\right]
\end{equation}
with $W=\text{diag}\left(\nicefrac{1}{\sum_{\xi}c_{i,\xi}\lambda_{\xi}}\right)$mimicking
a weighting matrix of a quadratic problem and $M=\text{diag}\left(\nicefrac{\lambda_{j}^{k}}{\eta_{j}}\right)$
the preconditioner.

Taking the Taylor expansion around the solution $\lambda^{*}$ of
equation \ref{eq:MLEM_PGD}, we find that 
\begin{equation}
\lambda_{j}^{k+1}-\lambda_{j}^{*}=I-\sum_{\xi}\left[\frac{1}{\eta_{j}}\sum_{i}c_{i,\xi}\frac{c_{i,j}\lambda_{j}^{k}\;y_{i}}{\left(\sum_{\zeta}c_{i\zeta}\lambda_{\zeta}^{k}\right)^{2}}\right]\left(\lambda_{\xi}^{k}-\lambda_{\xi}^{*}\right)\label{eq:convRate}
\end{equation}
where $I$ is the identity matrix. This equation closely mimic that
of an algorithm linearly converging. We will prove that we are able
to estimate an effective linear convergence rate $\alpha_{j}=\text{E}\left[\frac{\left\Vert \lambda_{j}^{k+1}-\lambda_{j}^{*}\right\Vert }{\left\Vert \lambda_{j}^{k}-\lambda_{j}^{*}\right\Vert }\right]$
for different kind of signals. To do this we need to compute $\text{E}\left\{ \sum_{\xi}c_{i,\xi}\left(\lambda_{\xi}^{k}-\lambda_{\xi}^{*}\right)\right\} $.
We also notice that in a Poisson problem $\text{E}\left(\frac{y_{i}}{\bar{y_{i}}^{2}}\right)=\frac{1}{\bar{y_{i}}}$
exactly, even at very low expected number of counts per bin. In MLEM,
after the first iteration, the total amount of activity estimated
in the FOV is constant. Therefore, in background regions, the expectation
value $\text{E}\left(\lambda_{\xi}^{k}-\lambda_{\xi}^{*}\right)=0$.
In pixels belonging to an uniform object of extension $d$ centered
in $j$ we can approximate 
\begin{equation}
\text{E}\left(\lambda_{\xi}^{k}-\lambda_{\xi}^{*}\right)=\left(\lambda_{j}^{k}-\lambda_{j}^{*}\right)\cos\left(\frac{\pi}{2}\,\frac{x}{d}\right)\;\;|x|<d\label{eq:BP_tails}
\end{equation}
and $0$ elsewhere, where $x$ is the distance between pixel $\xi$
and the pixel $j$. We used the cosine function extending outside
the object to approximate the low frequency effect of backprojection.
Under this assumption $\text{E}\left(\sum_{\xi}c_{i,\xi}\left(\lambda_{\xi}^{k}-\lambda_{\xi}^{*}\right)\right)=c_{i,j}\,\frac{2}{\pi}d\,\left(\lambda_{j}^{k}-\lambda_{j}^{*}\right)$,
and therefore into 
\[
\alpha_{j}=1-\frac{1}{\eta_{j}}\sum_{i}c_{i,j}\frac{c_{i,j}\lambda_{j}^{k}}{\sum_{\zeta}c_{i\zeta}\lambda_{\zeta}^{k}}\frac{2}{\pi}d=
\]
\begin{equation}
=1-\nicefrac{2}{\pi}\left\langle \frac{c_{i,j}\lambda_{j}^{k}}{\bar{y_{i}}}\right\rangle d\label{eq:convSphere}
\end{equation}
where brackets indicate the average over all the LORs contributing
to pixel $j$. We can easily observe that signal convergence depends
on its spatial extension and on its contribution to the total number
of counts in the LORs. If the image is composed by a background circle
of diameter $D$ with a smaller concentric circle of diameter $d$
and activity ratio $\beta$, in the central pixel $j$ we can further
simplify equation \ref{eq:convSphere} as
\begin{equation}
\alpha_{j}=1-\nicefrac{2}{\pi}\frac{d\beta}{D+(\beta-1)d}\label{eq:sphere_Simp}
\end{equation}
With TOF the same equation holds with $D_{eff}$ in place of $D$,
if $D_{eff}<D$. Equation \ref{eq:sphere_Simp} is an approximation
because, especially at low iterations and for small values of $\beta$,
the reconstructed contrast might be different than the true contrast
$\beta$. A few things can be noticed: 
\begin{enumerate}
\item With increasing signal to background contrast $\beta$, the convergence
is faster.
\item With increasing object size $d$, the convergence speed also increases.
\item If an object is totally cold, i.e. $\beta=0$, the convergence rate
approaches $1$. This makes the convergence of cold objects extremely
slow.
\item Without TOF the convergence speed decreases with the background diameter
$D$. With TOF, instead, convergence does not depend on $D$ (for
$D>D_{eff}$).
\item With TOF convergence depends on $D_{eff}$; if $d\approx D_{eff}$
the convergence is almost instantaneous.
\item For the full reconstruction of a circle, not only low frequency components
need to converge, but also high frequency components. Each frequency
converges according to the same equation by replacing $d$ with $\nicefrac{1}{f}$
.
\item The same model is able to describe the background noise convergence,
using $\beta=1$ and $d=\nicefrac{1}{f}$.
\end{enumerate}
From equation \ref{eq:sphere_Simp} it follows that signal convergence
is the geometric succession $\lambda_{j}^{k}=\lambda_{j}^{*}+\left(\lambda^{0}-\lambda_{j}^{*}\right)\left(\alpha_{j}\right)^{k}$.
As this equation is difficult to fit to data, we rewrite it introducing
an approximation
\begin{equation}
\frac{\lambda_{j}^{k+1}}{\lambda_{j}^{k}}=\frac{\lambda_{j}^{*}+\left(\lambda_{j}^{0}-\lambda_{j}^{*}\right)\left(\alpha_{j}\right)^{k+1}}{\lambda_{j}^{*}+\left(\lambda_{j}^{0}-\lambda_{j}^{*}\right)\left(\alpha_{j}\right)^{k}}=
\end{equation}
\[
=1+\frac{\left(\lambda_{j}^{0}-\lambda_{j}^{*}\right)\left[\left(\alpha_{j}\right)^{k+1}-\left(\alpha_{j}\right)^{k}\right]}{\lambda_{j}^{*}+\left(\lambda_{j}^{0}-\lambda_{j}^{*}\right)\left(\alpha_{j}\right)^{k}}
\]
\begin{equation}
\approx1+\frac{\lambda^{0}-\lambda_{j}^{*}}{\lambda_{j}^{*}}\left(\alpha_{j}\right)^{k+1}\left[1-\left(\alpha_{j}\right)^{-1}\right]\label{eq:logApprox}
\end{equation}
In this way we can perform the log-linear fit:
\begin{equation}
\ln\left(\frac{\lambda_{j}^{k+1}}{\lambda_{j}^{k}}-1\right)=\ln\varepsilon^{0}+\left(k+1\right)\ln\alpha_{j}+\ln\left(1-\nicefrac{1}{\alpha_{j}}\right)=\left(k+1\right)\ln\alpha_{j}+\text{const}\label{eq:semiLog-conv}
\end{equation}
with $\varepsilon^{0}=\nicefrac{\lambda_{j}^{0}-\lambda_{j}^{*}}{\lambda^{*}}$.
A second order expansion of the approximation introduced would result
in equation \ref{eq:semiLog-conv} becoming $=\text{const}+\ln\left(\left(\alpha_{j}\right)^{k+1}-\varepsilon^{0}\left(\alpha_{j}\right)^{2k+1}\right)$,
thus negligible after very few iterations. 

Equation \ref{eq:BP_tails} is an approximation of the backprojection
shape. Therefore in the next section, on top of validating this model
we will fit the proportionality constant, here found to be $\nicefrac{2}{\pi}$,
defining it as a parameter $\gamma$.

\subsection{Contribution of attenuation, random and scattered coincidences\label{subsec:Contribution-of-attenuation,}}

\subsubsection*{Attenuation}

Using a notation where $c_{i,j}$ represents the total probability
of a photon emitted from the pixel $j$ to be detected in detector
$i$, the attenuation is naturally accounted for in all the previous
equations, without any modification.

\subsubsection*{Random and scattered coincidences}

In presence of an expected rate of random coincidences $r$ and of
scattered coincidences $s$, equation (\ref{eq:MLEM_PGD}) becomes
\begin{equation}
\lambda_{j}^{k+1}=\lambda_{j}^{k}+\nicefrac{\lambda_{j}^{k}}{\eta_{j}}\sum_{i}c_{i,j}\frac{y_{i}-\sum_{\xi}c_{i,\xi}\lambda_{\xi}-r_{i}-s_{i}}{\sum_{\xi}c_{i,\xi}\lambda_{\xi}+r_{i}+s_{i}}
\end{equation}
 Therefore, equation (\ref{eq:convSphere}) still holds by considering
the contribution of scatter and random coincidences to $y_{i}$.

It can be seen that, the higher is the fraction of random and scattered
coincidences, the slower is the convergence rate. In TOF, random coincidences
are uniformly distributed over time bins. Pixel convergence is therefore
influenced not by the whole random amount, but only by the fraction
$\nicefrac{D_{eff}}{FOV}$ happening within $D_{eff}$. TOF indeed
greatly reduces the randoms impact on MLEM convergence. On the other
side, since the scatter coincidence profile over time bins follows
that of emission coincidences \cite{Watson2007}, the scatter impact
on convergence is not reduced with TOF. Furthermore, as the scatter
fraction greatly increases with object dimension, the previously observed
TOF convergence rate independence from background dimension is no
longer accurate.

\subsection{Noise convergence theory\label{subsec:Noise-convergence-theory}}

In order to study noise behavior over iterations, we revised the theory
of Barrett el al \cite{Barrett1994} . Briefly, the noise $\delta$
in the image is defined as a multiplicative factor, i.e. $\lambda^{k}=\bar{\lambda^{k}}\left(1+\delta^{k}\right)$,
where $\bar{\lambda^{k}}$ is the expected value of $\lambda$ at
iteration $k$. Basically $\delta$ is the relative error on a pixel
value. Working on the logarithms of the image and assuming that the
noise is low (i.e. $\delta\ll1$), $\log\lambda^{k}\approx\log\bar{\lambda^{k}}+\delta^{k}$.
Similarly, measured coincidences are defined as $y=\bar{y}+n$, where,
since photon detection is a Poisson process, $\text{E}\left(n\right)=0$
and the covariance matrix $\text{cov}\left(n\right)\equiv K=diag\left(\bar{y}\right)$.
It was then proved that $\delta^{k+1}=B^{k}n+\left[I-A^{k}\right]\delta^{k}$,
with $I$ the identity matrix and $B$ and $A$ two operators defined
as follow: 
\begin{equation}
B^{k}n=\frac{1}{\eta}H^{T}\left[\frac{n}{H\bar{\left[\lambda^{k}\right]}}\right]
\end{equation}
\begin{equation}
A^{k}\delta=\frac{1}{\eta}H^{T}\left[\frac{H\left[\bar{\lambda^{k}}\circ\delta\right]}{H\bar{\left[\lambda^{k}\right]}}\right]
\end{equation}
$B$ mostly involves a weighted backprojection, therefore it produces
a low frequency image. $A$ operates instead on images and can be
written in a form $H^{T}WH$, where $W$ is the diagonal statistical
weighting matrix already defined in \subsecref{Objects-convergence-theory}.
Li \cite{Li2011} later studied the same problem without assuming
$H\left[\bar{\lambda^{k}}\right]\approx y$ at all iterations, resulting
in an identical $B$ term and a small corrective factor for $A$.
We retain the approximation introduced by Barrett, as the impact of
the corrective term is small and otherwise we cannot derive explicit
expressions. 

It is instructive to note that, for a shift invariant system, when
all the projections have identical weighting, $A$ is the $\nicefrac{1}{r}$
lowpass filter and, therefore, $A^{-1}$ is the ramp filter. The introduction
of TOF modifies the ramp filter from $|r|$ to 
\begin{equation}
\frac{1}{\sqrt{2\pi}\sigma_{t}}e^{-\frac{r^{2}}{2\sigma^{2}}}|r|\label{eq:filter}
\end{equation}
 with $r$ the radial frequency and $\sigma_{t}$ the timing resolution.
Its inverse also is not anymore the $\nicefrac{1}{r}$ filter but
it is modified by letting all frequencies $r<\sigma_{t}$ passing
unaltered and penalizing less higher frequencies. 

Assuming that $\lambda^{0}$ is noiseless, we can express the noise
at each iteration as a function of sinogram noise $n$, using an operator
$U^{k}$ defined recursively as 
\begin{equation}
U^{k+1}=B^{k}+\left[I-A^{k}\right]U^{k}
\end{equation}
 so that $\delta^{k}=U^{k}n$ . The covariance matrix of $\delta$
results $K_{\delta}=UKU^{T}=U\,diag\left(H\bar{\lambda}\right)\,U^{T}$
\cite{Barrett1994}. Therefore the standard deviation on an individual
pixel $j$ is $std\left(\lambda_{j}\right)=\lambda_{j}\sqrt{K_{\delta_{j,j}}}$.

\subsubsection{Noise frequency analysis}

From the previous equation, the noise spatial frequency behavior over
iterations can be derived. As already done in \subsecref{Objects-convergence-theory}
we assume that $A$ and $B$ do not depend on the iterations, that
is $H\bar{\lambda^{k}}\approx\bar{y}$ . With this approximation the
recursive expression for $U^{k}$ becomes a geometric succession 
\begin{equation}
U^{k}=\sum_{l=0}^{k}\left(I-A\right)^{l}B=A^{-1}\left[I-\left(I-A\right)^{k}\right]B
\end{equation}
Expanding the power term, which is useful to compute a low iteration
limit,
\[
U^{k}=A^{-1}\left[a_{1}A-a_{2}A^{2}+a_{3}A^{3}-\dots\right]B
\]
\begin{equation}
=a_{1}B-a_{2}AB+a_{3}A^{2}B-\dots
\end{equation}
 where $a_{1},a_{2}\ldots a_{k}$ are the binomial coefficients for
the power $k$. Since $B$ is a backprojection and $A$ a lowpass
filter, low frequency noise components will converge much faster than
high frequencies, thus the noise at the first iterations will have
only low frequency components. Higher frequencies will be reconstructed
more slowly by the action of the $A$ operator. At late iterations,
$\left(I-A\right)^{k}\ll I$ , therefore $U^{k}\approx A^{-1}B$ and
noise will feature an enhancement of the high frequency components. 

With the introduction of TOF, the lowpass filter in $A$ has a higher
frequency cutoff, that moves to higher values with improving CTR.
Conversely, the high frequency enhancement provided by $A^{-1}$ becomes
less pronounced, as shown in equation \ref{eq:filter}. 

\subsubsection{Low iterations limit\label{subsec:Low-iterations-limit}}

At initial iterations, since the noise $\delta$ is small and with
zero average, we can assume $A\delta\ll Bn$ and $U^{k}\approx\sum_{i=1}^{k}B^{i}$.
If, as before, we further approximate $B$ as iteration independent,
then $U^{k}\approx kB$. In component notation 
\[
std(\lambda_{j})=\lambda_{j}\sqrt{K_{\delta_{j,j}}}=\lambda_{j}k\sqrt{\nicefrac{1}{\eta_{j}}\sum_{i}\nicefrac{c_{i,j}}{\bar{y_{i}}}K_{i,k}\sum_{l}\nicefrac{c_{l,j}}{\bar{y_{l}}}\nicefrac{1}{\eta_{j}}}=
\]
\begin{equation}
=\lambda_{j}k\sqrt{\nicefrac{1}{\eta_{j}^{2}}\sum_{i}\left(c_{i,j}\right)^{2}\frac{1}{\bar{y_{i}}}}
\end{equation}
where the last step follows from the definition of $K$. For the central
pixel of a uniform disk of diameter $D$, we can further simplify
the equation to 
\begin{equation}
std\left(\lambda_{j}\right)=k\lambda_{j}\left\langle \frac{1}{\sqrt{y_{i}}}\right\rangle \label{eq:noise slope}
\end{equation}
 where with the brackets we indicate the average over all the LORs
contributing to the $j-$th pixel.

At low iterations the pixel standard deviation linearly increases
with the iteration number. All the LORs intersecting the disk central
pixel have the same expected sinogram counts $y_{i}\propto D$, therefore
$std\left(\lambda_{j}\right)\propto k\lambda_{j}\frac{1}{\sqrt{D}}$.
With the introduction of TOF $y_{i}\propto D_{eff}$, resulting in
an increase of low iterations noise by a factor $\sqrt{D/D_{eff}}$.
At the same time, with improved CTR, the action of $A$ becomes negligible
at a much lower number of iterations, both because $\delta$ converges
faster and because the strength of the filter represented by $A$
is reduced. Therefore, the linearity range is smaller.

In the next section we will investigate the range where noise linearly
increases with iterations, and we will investigate the noise level
ratios between TOF and non-TOF reconstruction at full convergence.

\subsection{Random and scattered coincidences, and attenuation\label{subsec:Random-noise}}

In the original paper by Barrett \cite{Barrett1994}, the influence
of attenuation and of random and scattered coincidences was not analyzed.
As in section \ref{subsec:Contribution-of-attenuation,}, attenuation
is naturally modeled in the equations by using appropriate $c_{i,j}$
factors. Regarding random and scattered coincidences, it can be shown
that $A$ and $B$ are modified as follows:
\begin{equation}
B^{k}n=\frac{1}{\eta}H^{T}\left[\frac{n}{H\left[\bar{\lambda^{k}}\right]+s+r}\right]
\end{equation}
\begin{equation}
A^{k}\delta=\frac{1}{\eta}H^{T}\left[\frac{H\left[\bar{\lambda^{k}}\circ\delta\right]}{H\left[\bar{\lambda^{k}}\right]+s+r}\right]
\end{equation}
 In presence of random coincidences, noise convergence is slower.
Furthermore, the sinogram covariance matrix is modified as $K=diag\left(\bar{y}+s+r\right)$,
thus leading to a noise increment with increased random and scatter
coincidences rates.

\section{Simulations\label{sec:Simulations}}

\subsection{Simulator description}

Simulations were performed in an idealized setting. Matched distance-driven
projectors and backprojectors were used \cite{Manjeshwar2007}, simulating
a single slice of a scanner with a $829\,mm$ ring diameter, with
$4.3\,mm$ crystals. The same projectors were used for image reconstruction.
Images were generated at high resolution ($1.2\,mm$ pixel size),
smoothed with a $4.5\,mm\;FWHM$ Gaussian kernel, then forward projected.
When needed, Poisson noise was simulated. Sinograms were then reconstructed
in a $512\,mm$ FOV, using $2\,mm$ pixels and MLEM iterations, without
OS acceleration. Reconstructions were initialized with a unitary circle
as large as the FOV.

\subsection{Circle reconstruction speed}

\subsubsection{Methods}

In this section we simulated, in an idealized setting without attenuation
nor random or scattered coincidences, the convergence rate of circles
positioned in the center of a uniform background, with and without
TOF. Different circle diameters ($d=8,\,11,\,16,\,22\,mm$), contrasts
($\beta=1.7,\,2.0,\,2.4,\,3.3$) and background diameters ($D=492,\,465,\,437,\,410,\,383,\,355,\,328,\,300,\,273\,mm$)
were simulated. In TOF simulations, where we expect convergence rate
not to depend on the background diameter, we simulated only 3 diameters
($D=492,\,410,\,273\,mm$). CTR resolutions of $700;\,600;\,500\,ps\;FWHM$
were used, that correspond to a $D_{eff}=112,\,96,\,80\,mm$. No noise
was simulated.

The value of the central pixel of each image was analyzed with respect
to iterations and a log-linear plot was fitted to determine $\log\alpha_{j}$
according to equation \ref{eq:semiLog-conv}. We expect this relation
to be linear only in a limited range of iterations. In the earliest
ones, $y_{i}\not\approx\sum_{j}c_{i,j}\lambda_{j}$ and, furthermore,
the approximation introduced in equation (\ref{eq:logApprox}) to
obtain the lin-log relation does not hold yet. At late iterations,
low frequency components have already converged and high frequency
components dominate the update speed.

After determining $\alpha_{j}$ in all simulated conditions, we fitted
equation \ref{eq:convSphere} (in which we substituted $\nicefrac{2}{\pi}$with
$\gamma$), as we want to determine the proportionality constant independently
from all the assumptions.

\subsubsection{Results}

\begin{figure}
\subfloat[Activity value]{\includegraphics[width=0.49\columnwidth]{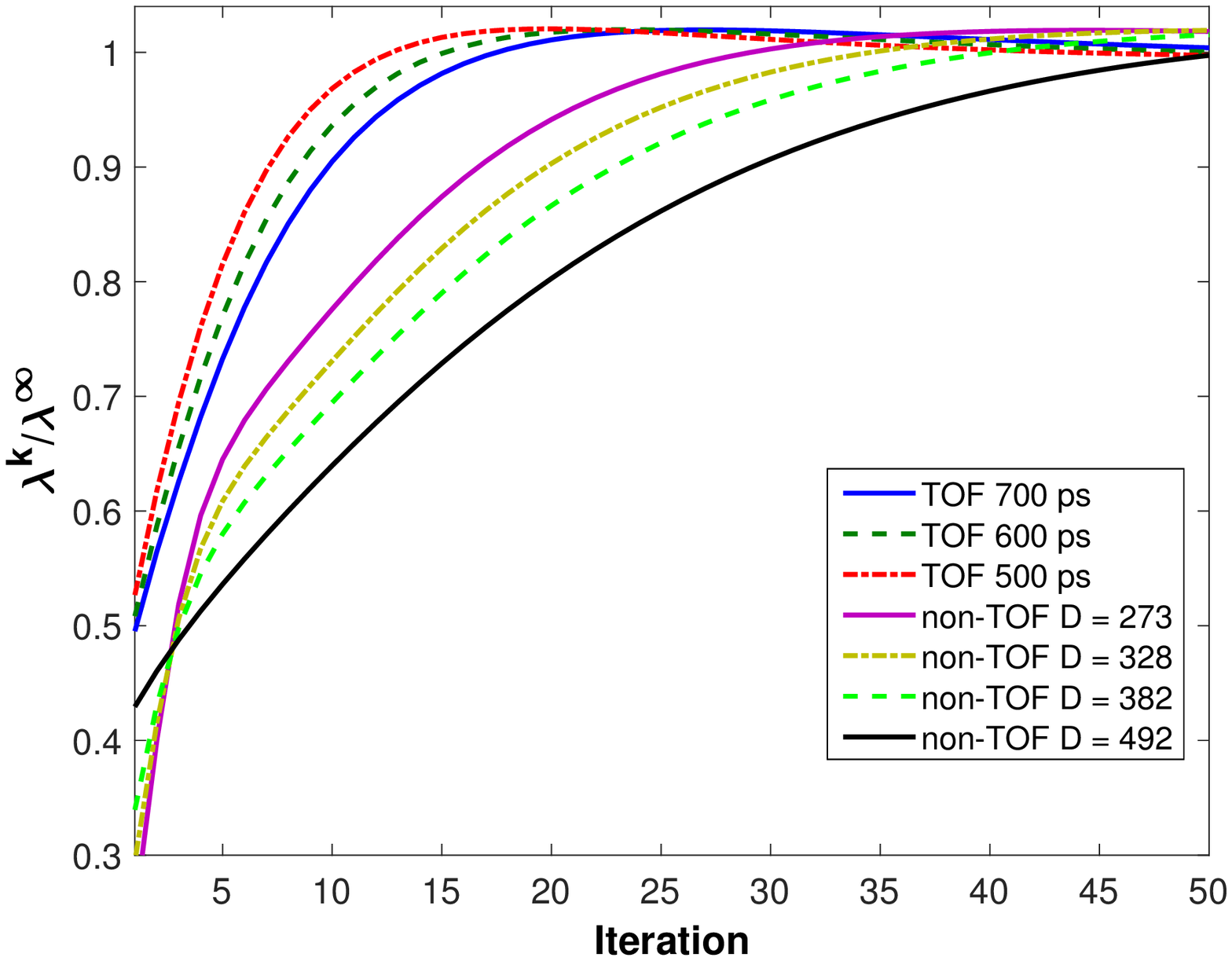}

}\subfloat[Linear-log plot of equation \ref{eq:semiLog-conv}]{\includegraphics[width=0.49\columnwidth]{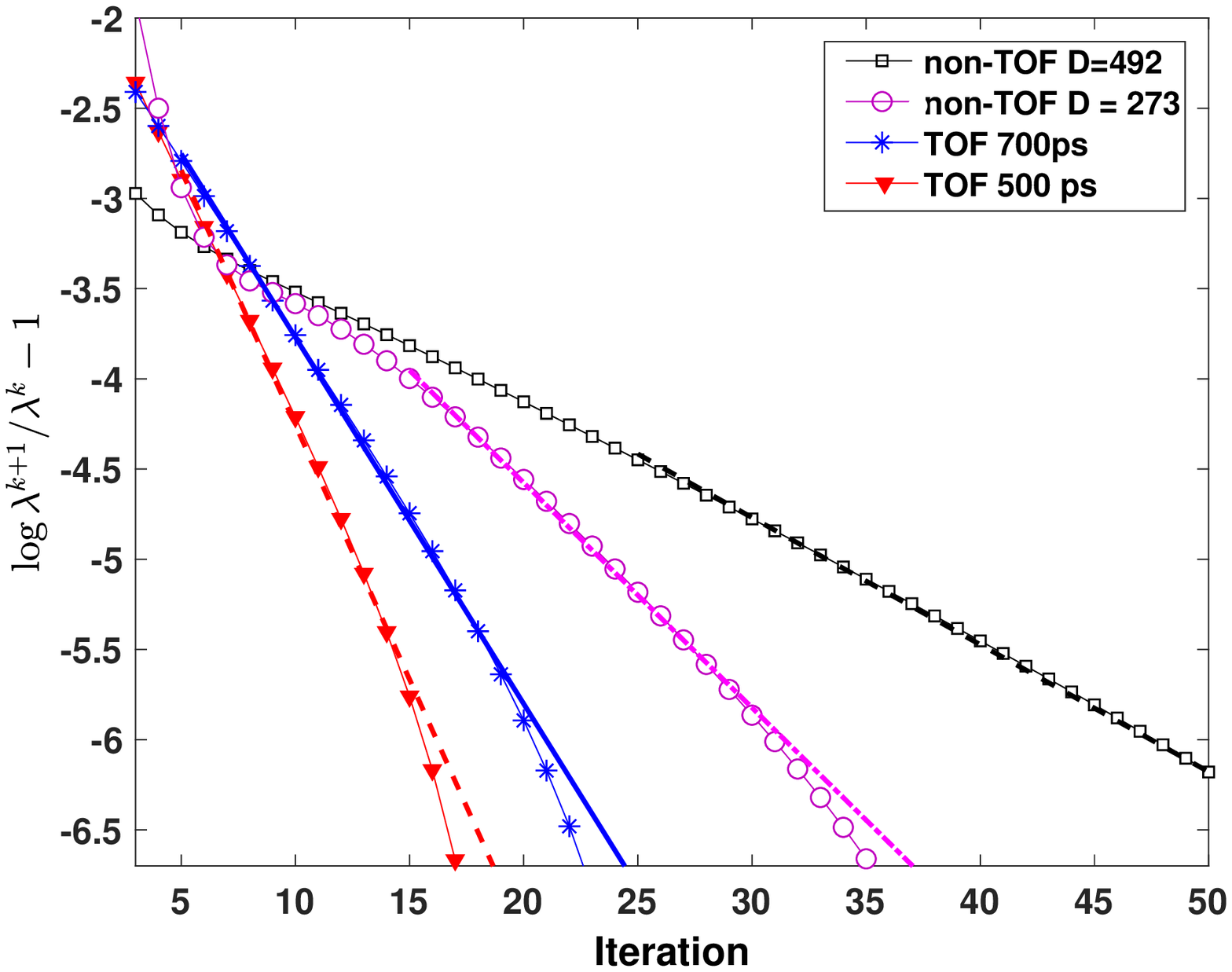}

}

\caption{\label{fig:Convergence-of-circles}Convergence of circles with contrast
$\beta=2.4$ and diameter $d=22\,mm$ for different background diameters
and CTR.}
\end{figure}
 \figref{Convergence-of-circles} shows the results of the simulations:
plots of activity values for representative circles (a), together
with the linear-log plots of equation \ref{eq:semiLog-conv} (b).
For TOF only $D=410\,mm$ curves are shown as they were perfectly
superimposed for all diameters. Plots of $\alpha_{j}$ as a function
of background diameter, contrast and object size are shown in supplementary
figure 1. For $\log\alpha_{j}\apprge0.30$, model fitting was difficult
due to the very high convergence speed. This prevented us to investigate
the model for higher contrasts and lower CTRs. Equation \ref{eq:sphere_Simp}
was found able to correctly describe the convergence rate for all
the diameters and contrasts investigated, with $\gamma=0.59\pm0.08$,
both for TOF and non-TOF recons. The discrepancy between predicted
and fitted $\gamma$ is acceptable considering the high number of
assumptions on $\alpha_{j}$ and $\lambda_{j}$ trends along a LOR.
As to TOF, simulations confirmed the independence of the convergence
speed from the background diameter. 

\subsection{Noise properties}

\subsubsection{Methods}

A uniform phantom of $410\,mm$ diameter was simulated and reconstructed
over a $512\,mm$ FOV using $2\,mm$ pixels. Images were forward projected
without and with TOF at CTR of $650,\,400,\,300\,,80\,ps$ FWHM that
correspond to a $D_{eff}=104,\,64,\,48\,,13\,mm$. Attenuation, random
and scatter were not simulated. A Poisson process was simulated to
achieve 12 independent noise realizations of the noiseless sinogram.
An identical total number of counts was simulated at different CTRs.
Images were reconstructed using MLEM without OS acceleration over
1000 iterations.

To quantify image noise, activity was sampled in 148 close but not
contiguous pixels at the image center. Noise was computed as the average
over the 148 pixels of the standard deviation among the 12 realizations.
A linear fit of noise vs the number of iterations was performed by
using only the first 5 iterations. The relation between the fit slope
and background diameter was compared to that predicted in equation
\ref{eq:noise slope}. The ratio between TOF and non-TOF noise at
convergence was compared to the predictions made in reference \cite{Vunckx2010}.

As previously stated, the convergence of each noise frequency $f$
should follow equation \ref{eq:convSphere} with $\beta=1$ and $d=\nicefrac{1}{f}$.
However, a linear-log fit cannot be performed, since noise in a uniform
disk simultaneously involves all possible frequencies. Therefore,
we analyzed the noise power spectrum at representative number of iterations,
and noise as a function of iterations on post-filtered images.

\subsubsection{Results}

\begin{figure}
\subfloat[\label{fig:Plots-of-noise}Plots of noise as a function of the iterations
on unfiltered images. Selected CTR shown for clarity]{\includegraphics[width=0.49\columnwidth]{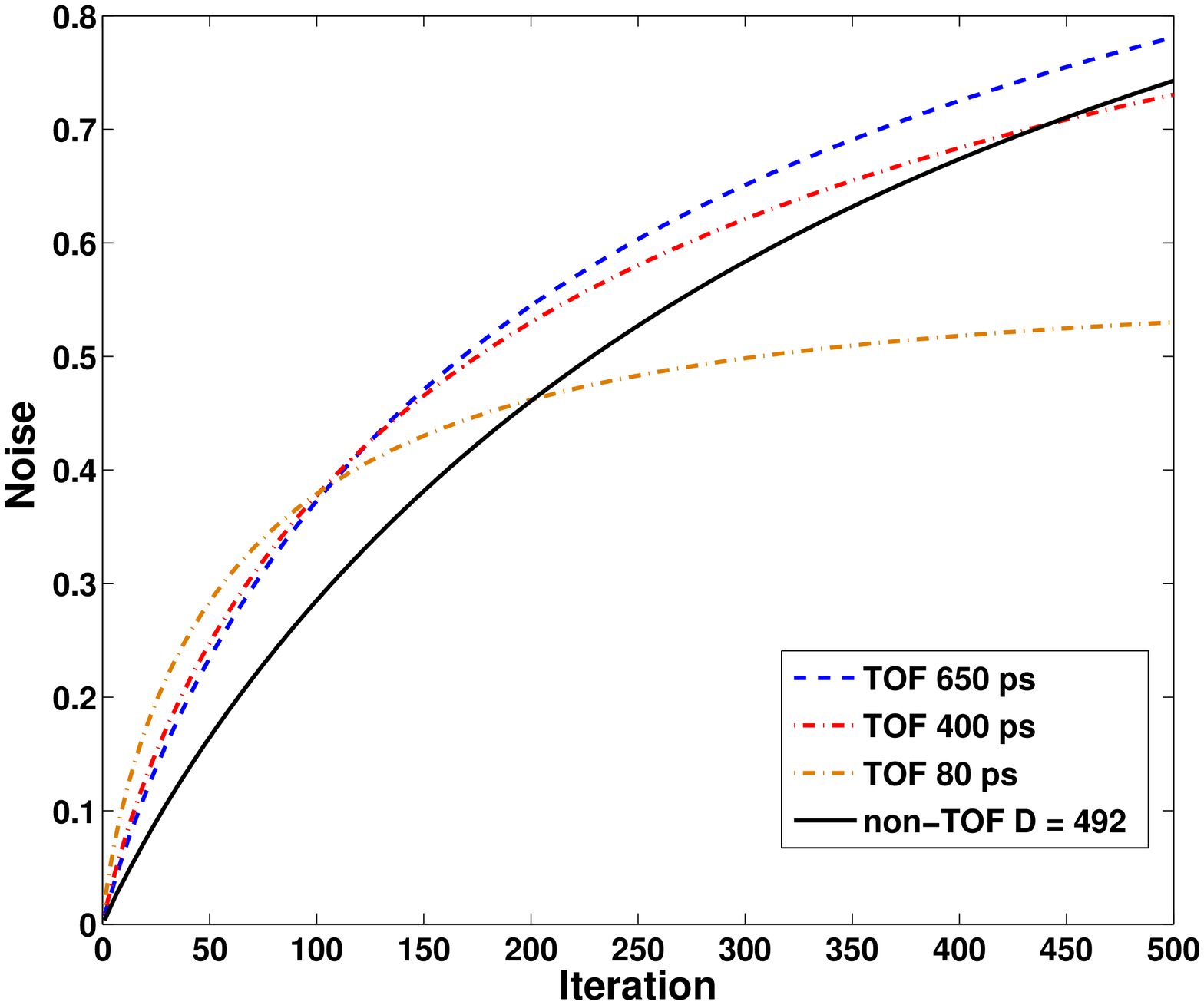}}\subfloat[\label{fig:Noise-smooth}Plot of noise as a function of iterations
for images post-smoothed with a $4.5\,mm$ Gaussian filter]{\includegraphics[width=0.49\columnwidth]{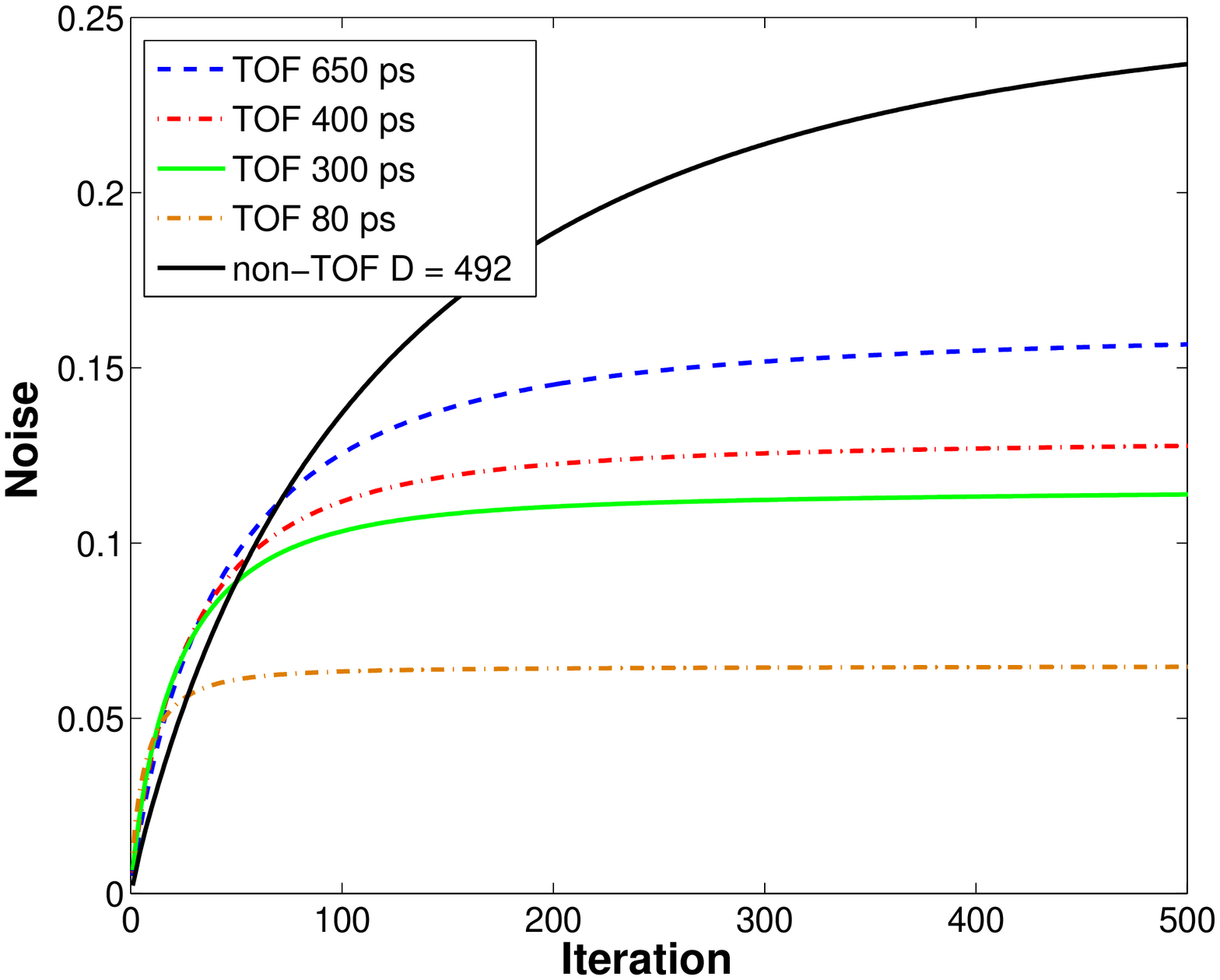}

} \caption{Noise as a function of iterations}
\end{figure}
 In \figref{Plots-of-noise} noise is shown as a function of iterations
for the first $500$ iterations. In the first iterations, shown in
supplementary figure 2, it can be properly described by the linear
model within a $25\%$ error up to $79$ iterations for non-TOF, to
$27$ iterations for TOF with $650\,ps$ CTR, $16$ iterations for
$400\,ps$ and $8$ iterations for $80\,ps$. The slopes follow the
$\nicefrac{1}{\sqrt{D_{eff}}}$ trend as predicted in section \ref{subsec:Low-iterations-limit}.
With TOF at $650$ or $400\,ps$ is higher than in non-TOF up to $\approx500$
iterations. Up to $1000$ iterations MLEM has not reached convergence
yet. Ratios between TOF and non-TOF noise at convergence ($6000$
iterations) corresponds to predictions in reference \cite{Vunckx2010}
(results not shown). In supplementary figure 3 we show the noise power
spectrum at different number of iterations for non-TOF and TOF with
$400\,ps$ CTR. At low iterations, the noise power spectrum resembles
the $\nicefrac{1}{r}$ trend, while at convergence it approximates
the ramp. This was expected under the considerations introduced in
(\ref{subsec:Low-iterations-limit}). The low frequency noise power
is lower with TOF than without it. As to higher frequencies, without
TOF, they are still far from convergence at $1000$ iterations, while
with TOF the usual shape of the ramp filter can be appreciated. In
\figref{Noise-smooth} we show noise as a function of iterations,
on images post-filtered with a $4.5\,mm\,FWHM$ Gaussian kernel. Noise
converges much quicker than without any smoothing, nonetheless, TOF
reconstructions with CTR of $400\,ps$ still have noise higher than
non-TOF up to $\approx60$ iterations. 

\section{Phantom measurements\label{sec:Phantom-measurements}}

\subsection{Methods}

\subsubsection{Phantom preparation and acquisition}

Two $22\,cm$ diameter Derenzo phantoms were filled with $^{18}F-FDG$.
The first phantom has the cold insert and was filled with $15.5\;kBq/cc$
of activity, calibrated at the time of the first scan. In the uniform
region, six cold spheres of $8,\,11,\,14,\,17,\,24$ and $30\,mm$
diameters were positioned. The second phantom has the hot insert and
was filled with $15.4\;kBq/cc$ of activity in the background region.
It also featured 6 hot spheres of $7,\,12,\,15,\,16,\,19$ and $24\,mm$
diameter, filled with a $4.48:1$ activity ratio vs the background.
The phantoms were acquired both individually (``Single'' configuration)
and side by side (``Double'' configuration), to simulate different
object dimensions. In the double configuration, the active area has
a major axis of about $44\,cm$ and a minor axis of $22\,cm$, which
is close to the dimension of a typical patient. 

To investigate the effect of TOF at different CTRs, the phantoms were
scanned in two different PET tomographs: a General Electrics Medical
Systems SIGNA PET/MR, with $400\,ps$ CTR and a GEMS Discovery D690
with $650\,ps$ CTR. To better investigate the random coincidences
effect, the double configuration was scanned twice in the SIGNA: the
first time immediately after preparation with a true to random coincidences
ratio $\approx1$ (``HCR'' high count rate configuration), and the
second time $2\,h\,21\,min$ later, with a true to random coincidences
ratio $\approx5:1$ (``LCR'' low count rate configuration). Each
configuration was acquired for 10 minutes in list mode. Data were
then unlisted into 5 frames of 2 minutes to simulate multiple noise
realizations. 

\subsubsection{Image reconstruction}

To analyze early convergence, few iterations of MLEM without OS acceleration
were performed (SIGNA: TOF: 20, non-TOF: 40. D690: TOF: 30, non-TOF:
60). To study the convergence of high frequencies and noise with and
without TOF, a second reconstruction was performed with a full 3D-OSEM
algorithm using 16 subsets and 80 iterations (1280 updates or MLEM
equivalent iterations). Images were reconstructed using a $256\times256$
image matrix, on the maximum scanner FOV($70\,cm$ D690; $60\,cm$
SIGNA) and were post-filtered with a $4.5\,mm\;FWHM$ in-plane Gaussian
filter.

\subsubsection{Image analysis}

To study the signal convergence rate, the activity measured in the
central pixel of each sphere was fitted to equation \ref{eq:semiLog-conv}
. To quantify image noise, activity was sampled in 320 close but not
contiguous pixels of both phantoms. Noise was computed as the average
over the 320 pixels of the standard deviation among the 5 noise realizations. 

\subsubsection{Estimation of random and scatter coincidences}

In these phantom experiments, high fractions of random and scatter
coincidences are present. To fit equation \ref{eq:convSphere}, we
need to estimate the ratio between true counts and random and scattered
counts in each sinogram bin. We thus forward projected the reconstructed
image and multiplied the resulting sinogram by the attenuation and
the normalization sinogram. We then used scatter and random coincidence
sinograms to compute count ratios.

\subsection{Results}

\subsubsection{Sphere convergence}

\begin{figure}
\subfloat[Convergence of activity]{\includegraphics[width=0.45\columnwidth]{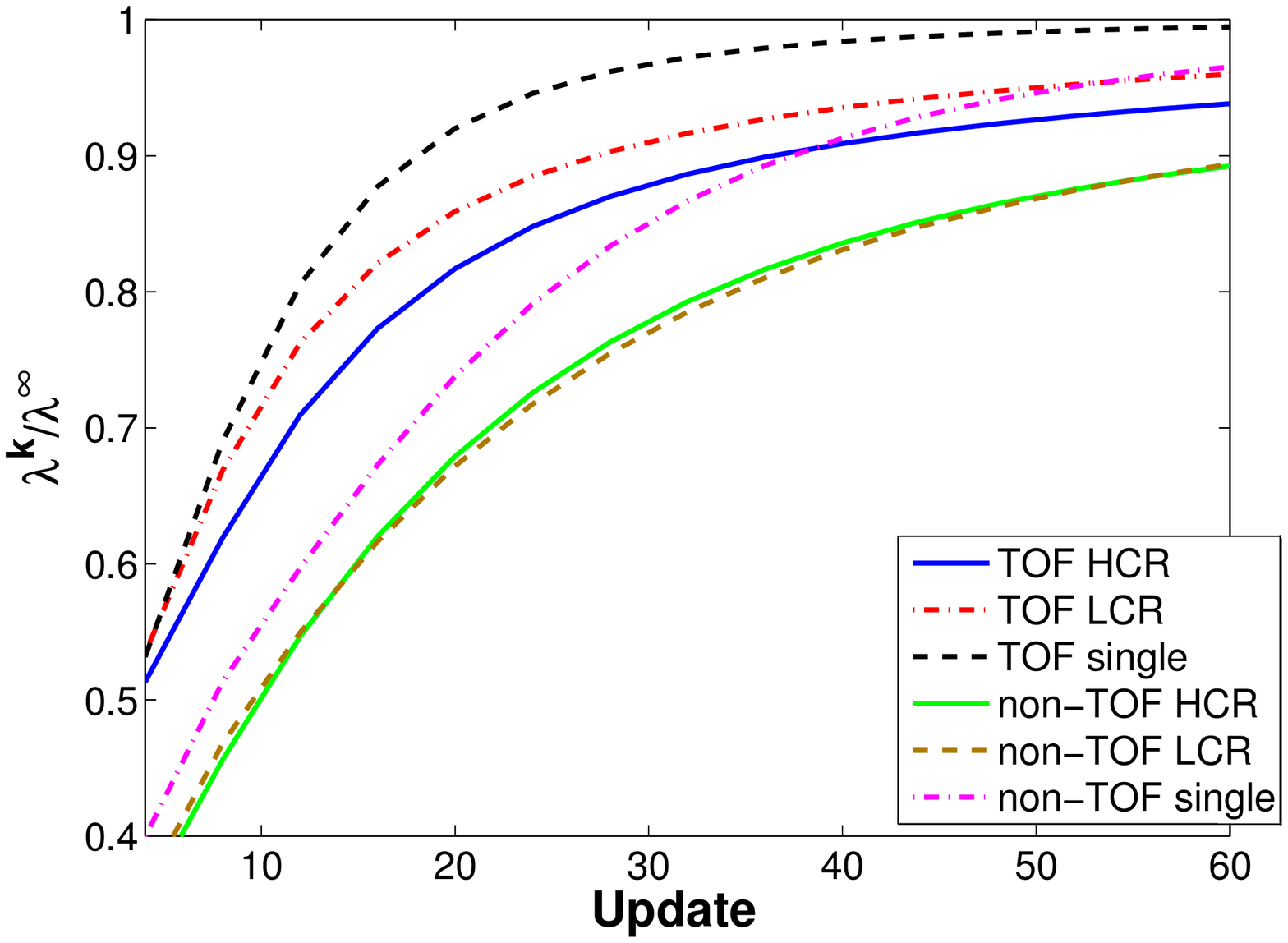}}\subfloat[Linear-log plots of equation \ref{eq:semiLog-conv}. Only 4 acquisitions
shown for clarity ]{\includegraphics[width=0.47\columnwidth]{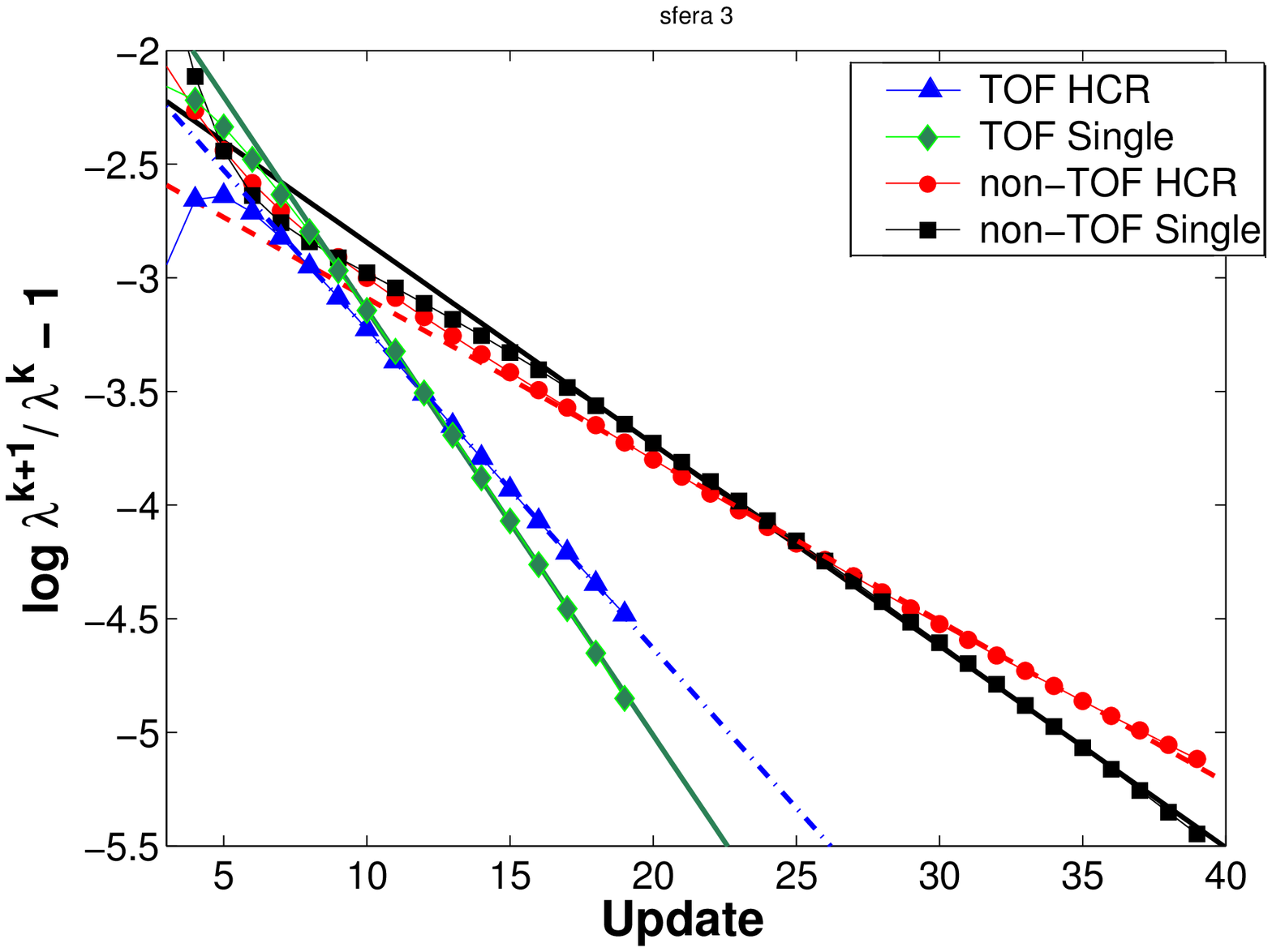}}

\caption{\label{fig:Convergence-signa}Convergence analysis of the $15\,mm$
sphere acquired on the SIGNA}
\end{figure}

The convergence of a representative sphere, together with its linear-log
plots are shown in \figref{Convergence-signa}. The convergence was
well described by the the linear fit. Nonetheless, since spheres are
not in the background center and due to the presence of scatter and
random coincidences, the coefficient $\alpha_{j}$ can not be simply
described in terms of contrast and dimension like in equation \ref{eq:sphere_Simp},
but the whole equation \ref{eq:convSphere} must be used. In \tabref{Alpha_signa}
we show fitted values of $\alpha_{j}$ for the $15\,mm$ sphere for
the different configurations in the two PET scanners.
 
\begin{table}
\begin{tabular}{|>{\centering}p{0.19\columnwidth}|>{\centering}p{0.15\columnwidth}|>{\centering}p{0.15\columnwidth}|>{\centering}m{0.13\columnwidth}|>{\centering}m{0.13\columnwidth}|}
\hline 
Acquisition & non-TOF D690 & non-TOF SIGNA & TOF D690 & TOF SIGNA \tabularnewline
\hline 
\hline 
Single & 0.099 & 0.085 & 0.115 & 0.171\tabularnewline
\hline 
Double LCR & 0.067 & 0.068 & 0.136 & 0.143\tabularnewline
\hline 
Double HCR &  & 0.077 &  & 0.131\tabularnewline
\hline 
\end{tabular}

\caption{\label{tab:Alpha_signa}Values of $\alpha_{j}$ for the $15\,mm$
sphere}
\end{table}
Without TOF, D690 and the SIGNA scanner values are similar. When TOF
is introduced, SIGNA has a higher convergence speed, thanks to its
better CTR.

\subsubsection{Noise proprieties}

\begin{figure}
\subfloat[Noise in the first iterations. LCR not shown for clarity ]{\includegraphics[width=0.48\columnwidth]{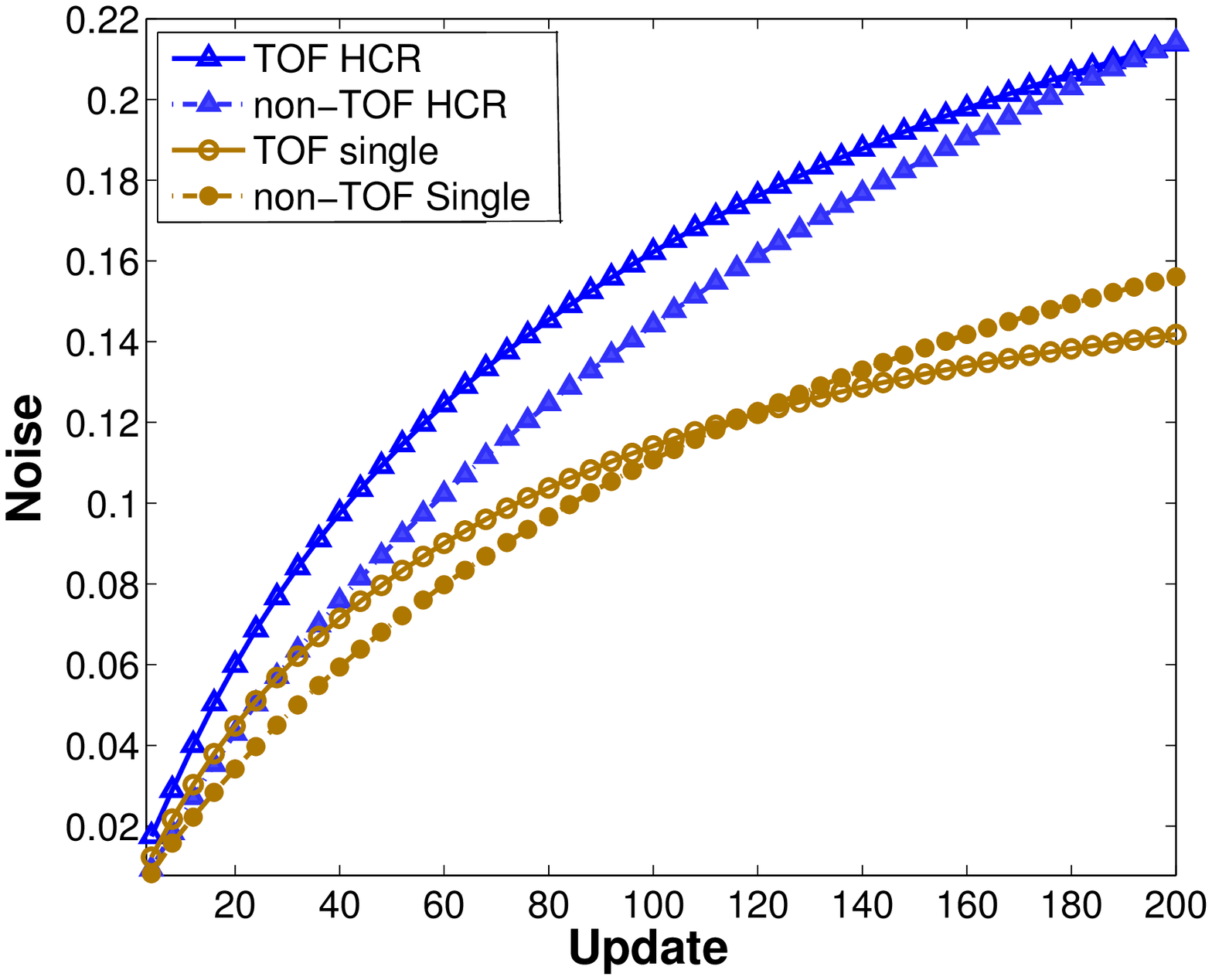}

}\subfloat[Noise trend at all iterations reconstructed]{\includegraphics[width=0.48\columnwidth]{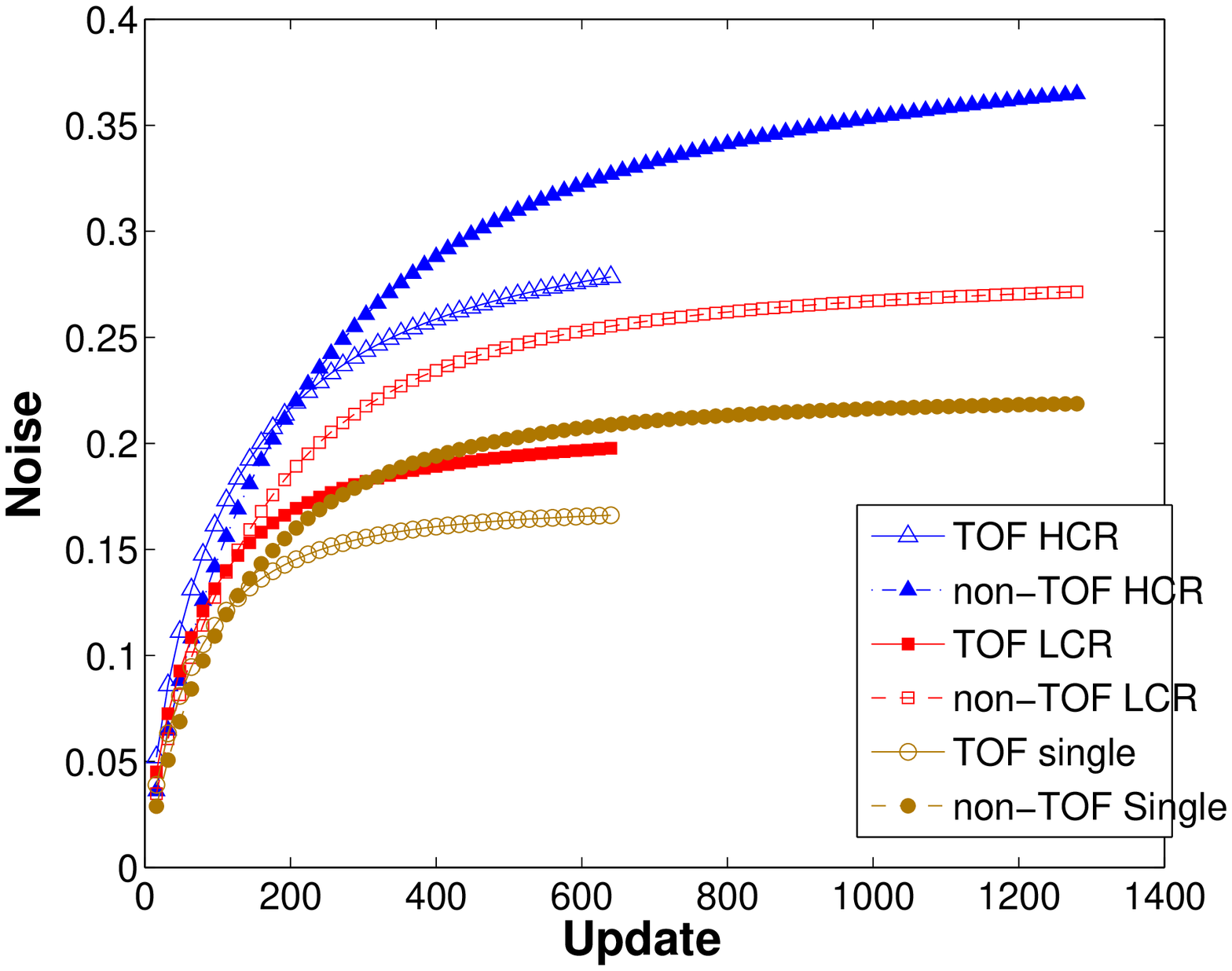}

}

\caption{\label{fig:Noise-signa}Noise for TOF and non-TOF measured in the
phantoms acquired in the SIGNA scanner.}
\end{figure}

In supplementary figure 4 we show the noise trends as a function of
iterations for D690 acquisitions. In figure \ref{fig:Noise-signa}
we show the same trends for the SIGNA scanner acquisitions. Noise
is higher with TOF than without it up to $200$ updates in the HCR
configuration and up to $120$ updates in the Single and the LCR configurations.
Curves are more detached than those obtained on the D690 scanner,
because of the lower CTR.
 At $1280$ updates noise in non-TOF has not converged yet. In table
\ref{tab:Noise-increase-table} we report the noise increase from
non-TOF to TOF reconstructions at 48 updates, a common clinical setting.
In the single configuration, in the D690 scanner almost no differences
are observed, as the CTR is not significantly smaller than the phantom
diameter. On the SIGNA scanner the increase is larger than in the
D690 scanner.
\begin{table}
\begin{tabular}{|c|c|c|}
\hline 
 & SIGNA & D690\tabularnewline
\hline 
\hline 
 & {[}\%{]} & {[}\%{]}\tabularnewline
\hline 
Single & +16.9 & +4.2\tabularnewline
\hline 
Double LCR & +13 & +10.0\tabularnewline
\hline 
Double HCR & +25.6 & \tabularnewline
\hline 
\end{tabular}

\caption{\label{tab:Noise-increase-table}Noise increase at 48 updates between
TOF and non-TOF reconstructions}
\end{table}

\subsubsection{Impact of scatter and random coincidences}

Results in \tabref{Noise-increase-table} show that the noise increase
from non-TOF to TOF is higher in HCR configuration than in the LCR
one. This is in accordance with what obtained in \subsecref{Random-noise}.
In supplementary figure 5 we show, for the LCR SIGNA acquisitions,
the estimated trues, scatter and random coincidences along the TOF
dimension for two orthogonal projections along the major and the minor
axis and of the phantom. As expected, the scatter distribution follows
the trues coincidences distribution, thereby limiting the improvements
achievable by TOF.

\section{Proposed early stopping criteria\label{sec:Proposed-early-stopping}}

Currently, in non-TOF reconstructions, it is customary to estimate
the optimal trade-off between signal recovery and noise on image quality
phantoms. Such phantoms are generally small in diameter compared to
patients and signal recovery is measured in a relatively high contrast
setup (e.g.: $10:1$). As we have shown, if the stopping criteria
are optimized on a high signal contrast and with a small diameter
background, when they are applied to patient areas like the abdomen,
characterized by a large diameter and by the presence of high uptake
organs in the background, a serious risk of signal under-recovery
may appear. Nonetheless, the widespread adoption of early stopping
in the last decades has shown that such criteria are effective enough
to allow a robust diagnosis. 

We here propose a stopping rule definition method for TOF to maximally
exploit TOF potentialities for a given combination of PET scanner,
injected dose and acquisition duration. First, a non-TOF reconstruction
stopping criteria $N_{it}$ has to be established according to the
current standard, using an image quality phantom, possibly with a
lower contrast (e.g $4:1$). Then, the optimal stopping criteria for
TOF can be obtained as $N_{TOF}=N_{it}\times\nicefrac{D_{eff}}{20\,cm}$
. On $20\,cm$ objects, this stopping criteria would provide the noise
reduction theoretically expected for TOF \cite{Vunckx2010} and a
signal recovery comparable to that achievable with non-TOF reconstruction.
For objects smaller than $20\,cm$, a slight under-recovery with respect
to non-TOF reconstructions would be present, but signal recovery in
small objects is generally not an issue. On objects larger than $20\,cm$
this criteria maintains the signal recovery determined on the image
quality phantom and still provide a noise reduction, even if by a
factor smaller than $\sqrt{\nicefrac{D_{eff}}{D}}$. 

\begin{figure}
\subfloat[First slice]{\includegraphics[width=0.35\columnwidth]{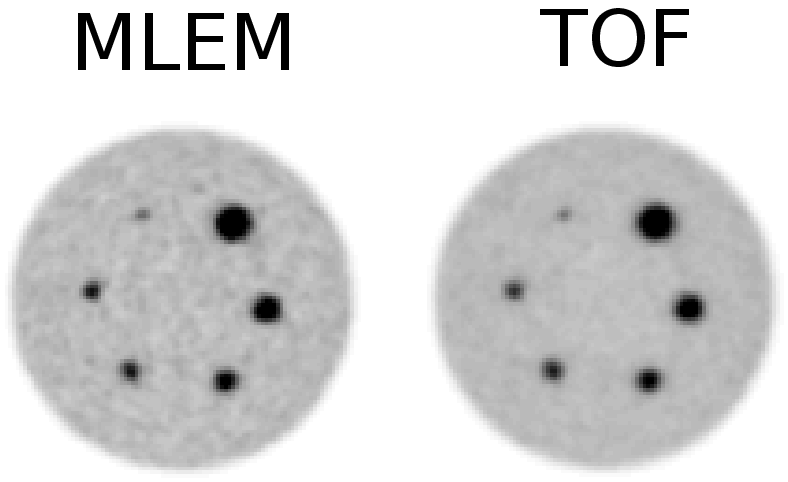}

}\subfloat[Second slice]{\includegraphics[width=0.35\columnwidth]{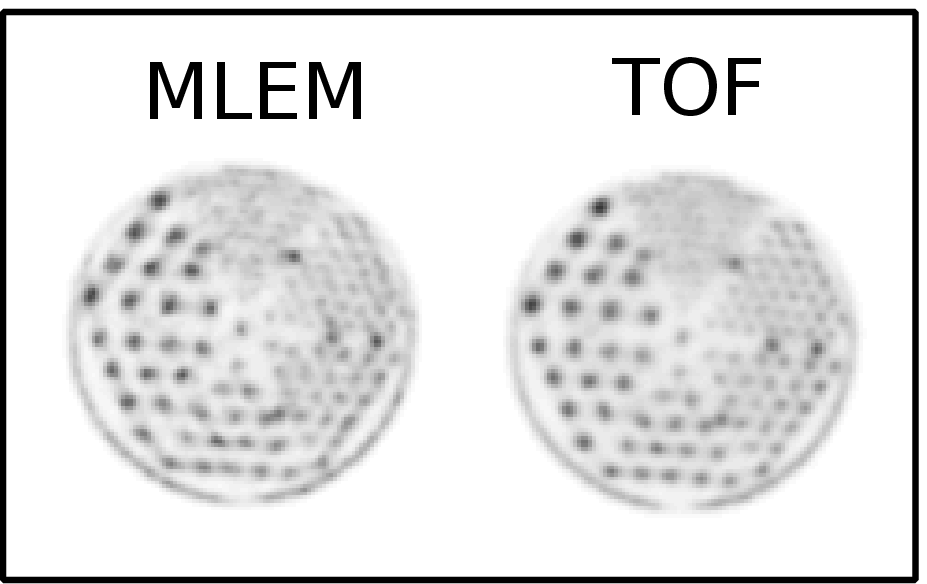}

}\caption{\label{fig:Hot-single_matched}Two slices of the phantom in the single
configurations, reconstructed using the proposed stopping criteria}
\end{figure}

\begin{figure}
\subfloat[First slice]{\includegraphics[width=0.35\columnwidth]{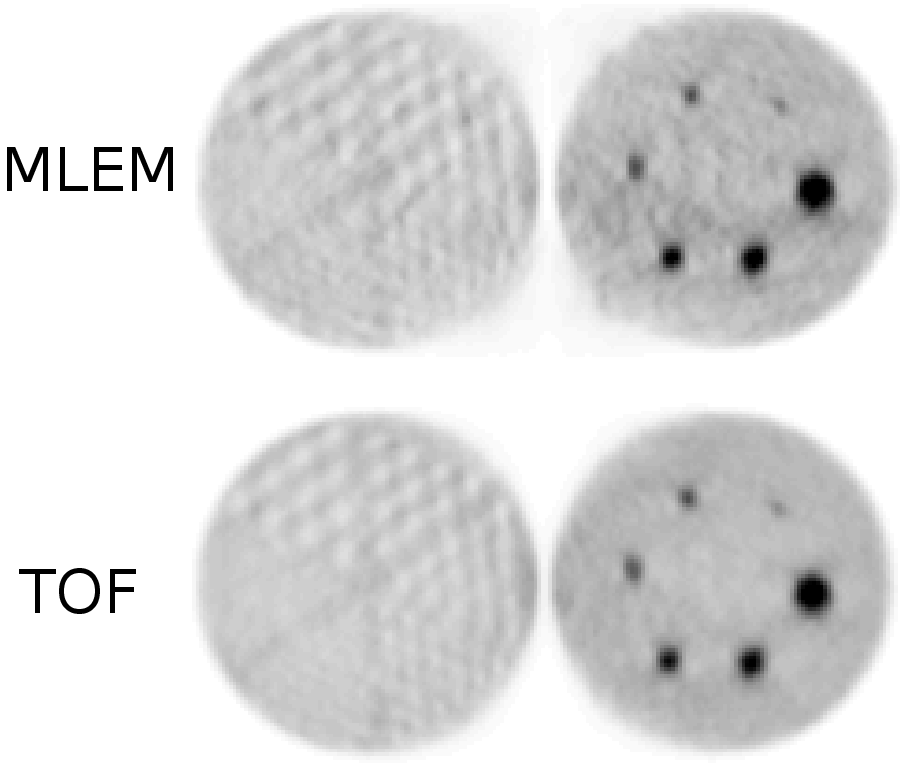}

}\subfloat[Second slice]{\includegraphics[width=0.35\columnwidth]{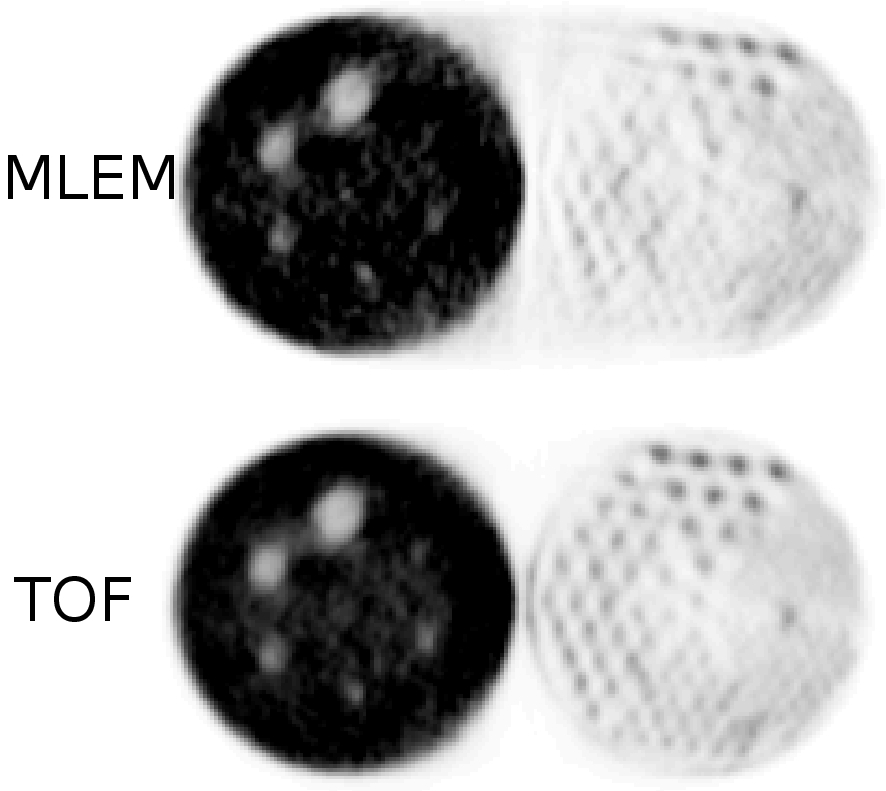}

}

\caption{\label{fig:Double-matched}Double phantom reconstructed with the proposed
criteria}
\end{figure}

The effects of this criteria can be qualitatively appreciated by analyzing
results obtained on the SIGNA scanner. On the Single configuration
of the hot phantom, we estimated 48 updates to be a reasonable stopping
criteria for non-TOF reconstruction. According to the proposed criteria,
we selected 16 updates for TOF reconstruction. In figure \ref{fig:Hot-single_matched},
two slices of the phantom obtained with 48 updates of non-TOF and
16 updates of TOF are shown. In the slice containing hot spheres,
detectability is visually similar with TOF, while noise is markedly
reduced. In the slice with the Jaszczak insert, TOF provides a visually
better image quality, despite the lower number of iterations. 
 In figure \ref{fig:Double-matched}, images of the Double configuration
phantom obtained with the same reconstruction strategies are shown.
In the slice containing the hot spheres and the cold Jaszczak insert,
TOF shows slightly better hot sphere contrast at much lower noise,
and a higher cold contrast. In the slice containing the cold spheres
and the hot Jaszczak insert, TOF markedly improves image quality respect
to non-TOF, that at 48 updates, is not able to recover details at
the FOV center yet. 

\section{Discussion}

In this paper we revised PET MLEM convergence properties, with the
aim to investigate TOF influence on signal recovery and noise amplification
on early stopped reconstructions, as these are still the most widely
used in clinical routine. In the theoretical analysis it was found
that signal convergence speed increases with signal contrast and dimension,
while it decreases with background diameter. TOF, in an ideal condition
without random and scatter coincidences, is able to eliminate the
dependence of signal convergence on the background diameter, thus
making the use of early stopped reconstructions more robust to variations
in patient size and background composition. Random and scatter coincidences
generally decrease the convergence speed. Importantly, TOF is able
to significantly reduce the impact of random coincidences on convergence;
scatter coincidences, instead, increase with patient size and, due
to their distribution, their impact is not greatly reduced by TOF.
All these findings were theoretically proved and confirmed by simulations
and phantom experiments.

As to noise behavior with iterations, higher frequencies converge
more slowly than lower ones, thanks to the low-pass effect of the
backprojection forward-projection sequence, base of the MLEM algorithm.
TOF reduces this low-pass filtering effect, thus making noise convergence
faster. While at convergence TOF reduces noise respect to non-TOF
by a factor $\sqrt{\frac{D_{eff}}{D}}$, if matched iterations are
used in early iterations TOF actually increases noise by the inverse
of this factor, thus resulting in sub-optimal exploitation of TOF
gains. In simulations, we found that TOF with $400\,ps$ CTR had noise
higher than non-TOF up to$\approx60$ iterations. In phantom experiments,
this limit is shifted to higher iteration values, because of the influence
of random and scattered coincidences on the convergence speed. On
the SIGNA scanner, the crossing point was obtained at $120$ iterations
for the Single configuration ($22\,cm$ axis) and at $200$ iterations
for the Double configuration ($44\,cm$ major axis). In the HCR configuration,
characterized by a high number of random and scattered events, at
$50$ iterations noise was increased by TOF of $25\%$. For this reason,
until regularization algorithms become routine in clinical applications,
we suggest a criteria to properly reduce the iteration number when
using TOF, to optimize TOF benefits in early-stopped MLEM. Stopping
at $\nicefrac{D_{eff}}{20\,cm}$ of the iterations used without TOF,
provides matched signal recovery and noise reduction for small objects,
and greatly improves recovery for large objects.

The novelty of this work lies in the theoretical derivation of the
convergence of signal and noise in MLEM, with and without TOF. In
non-TOF systems, using a fixed number of iterations results in approximately
constant levels of noise for all body shapes and regions, due to the
proprieties highlighted in \subsecref{Noise-convergence-theory}.
The signal recovery instead is strongly influenced by all these factors,
therefore the recovery factors measured in phantoms might severely
overestimate the recovery factors for low contrast objects in large
backgrounds (typically liver lesions, especially in obese patients).
The proposed criteria instead, guarantees that when TOF is used approximately
the same contrast recovery measured on the phantom used to choose
the stopping criteria is obtained in all imaging conditions, achieving
therefore both a noise reduction and also a more robust quantification.

\section{Conclusion}

The convergence proprieties of signal and noise in MLEM were revised,
taking into account the effect of TOF. Previous empirical findings
of increased convergence speed both of signal and noise were confirmed.
Finally, we introduced a way to determine using phantoms a stopping
criterion that guarantees background-independent levels of contrast
recovery, that can be determined a priori on phantoms.

\bibliographystyle{unsrt}
\addcontentsline{toc}{section}{\refname}\bibliography{xampl,library}

\end{document}